\documentclass[aps,prd,reprint,superscriptaddress,showpacs,nofootinbib]{revtex4-2}
\usepackage{amsmath}
\usepackage{amsfonts}
\usepackage{amssymb}
\usepackage{hyperref}
\usepackage{hyperref}
\usepackage{xcolor}
\usepackage{graphicx}
\usepackage{caption}        
\usepackage{subcaption}      
\usepackage{ragged2e}
\hypersetup{colorlinks=true, linkcolor=blue, citecolor=blue, urlcolor=blue}

\begin{document}
	\title{Enhanced Sensitivity and Noise Resilience in Two-Qubit Quantum Magnetometers}
	
	\author{S.~Nohekhan Shishavan}
	\affiliation{Department of Physics, Urmia University}
	
	\author{K. Aghayar Gharehbagh
		\thanks{Corresponding author: k.aghayar@urmia.ac.ir}}
	\email{k.aghayar@urmia.ac.ir} 
	\affiliation{Department of Physics, Urmia University}
	
	\author{H. Sedgi Gamichi}
	\affiliation{Department of Physics, Urmia University}
	
	\date{\today}
	\begin{abstract}
		We present a novel two-qubit quantum magnetometer Hamiltonian optimized for enhanced sensitivity and noise resilience. Compared to existing models, our formulation offers advantages in accuracy, robustness against noise, and entanglement dynamics. Using analytical methods, we derive the Quantum Fisher Information (QFI) and the Signal-to-Noise Ratio (SNR), highlighting its practical viability for magnetic field sensing. Our approach bridges theoretical insights with real-world applicability. We further analyze the performance of the magnetometer with a different initial entangled state, revealing the benefits of entanglement for sensitivity. A comparative analysis with leading research in the field underscores the advancements offered by our proposed design. Finally, we discuss the limitations of our current study and suggest potential avenues for future research.
	\end{abstract}
	
	\maketitle

\section{Introduction}

Quantum magnetometry, a prominent area within quantum sensing, has witnessed substantial progress in recent years, offering the potential for sensitivity and precision exceeding that of classical counterparts. The fundamental advantage of quantum magnetometry lies in its exploitation of quantum resources such as superposition, interference, and entanglement to achieve high-resolution measurements of magnetic fields. These quantum phenomena enable sensitivities that can scale more favorably with the number of sensing units, like qubits, compared to traditional sensors. Furthermore, quantum magnetometers are often microscopic and operate with minimal invasiveness, making them suitable for a wide array of applications, ranging from fundamental physics investigations to medical diagnostics and industrial quality control \cite{degen2017quantum}.

In the pursuit of enhanced sensitivity and robustness in quantum magnetometry, multi-qubit systems have emerged as promising candidates, primarily due to their unique entanglement properties \cite{huang2024entanglement}. Increasing the number of qubits generally leads to improved sensitivity by enhancing the signal-to-noise ratio during the sensing process \cite{rovny2025multiqubit}. This occurs because more qubits can accumulate a greater amount of phase information that is dependent on the magnetic field being measured. However, it is also noted that increasing the number of qubits might lead to a decrease in spatial resolution, potentially due to the larger physical space required to accommodate them \cite{zhao2012decoherence}. 

Various quantum magnetometer technologies currently exist, each with its own set of operating principles, sensitivity levels, and limitations \cite{barry2020sensitivity}. Examples include Superconducting Quantum Interference Devices (SQUIDs), which are known for their high sensitivity \cite{budker2007optical}; optically pumped atomic magnetometers, which also offer exceptional sensitivity \cite{budker2007optical}; and magnetometers based on Nitrogen-Vacancy (NV) centers in diamond, which are promising for nanoscale sensing \cite{wolf2015subpicotesla, rondin2014magnetometry}. The choice of magnetometer technology is often dictated by the specific requirements of the application, such as the desired sensitivity, the frequency range of the magnetic field, and the environmental conditions under which the sensor must operate \cite{budker2007optical}.

This study introduces a novel two-qubit quantum magnetometer Hamiltonian specifically designed to achieve enhanced sensitivity and resilience to noise. The proposed Hamiltonian incorporates several key components, including the Zeeman interaction, spin-spin interactions (both longitudinal and transverse), terms modeling environmental dephasing, and time-dependent driving terms. The design philosophy behind this Hamiltonian is to strike a balance between theoretical sophistication and practical applicability by including terms that not only capture the essential physics of the system but also model realistic experimental conditions and allow for precise control over the qubits. By leveraging these innovative features, the aim is to demonstrate the superiority of this approach in terms of sensitivity, noise resistance, and experimental feasibility. Following the introduction in this section, Section II will delve into the methodology, providing the Dyson series, the conditions for its convergence, and the implementation of the quantum sensing protocol. Section III will present and discuss the results obtained from both analytical calculations and numerical analyses. Finally, concluding remarks, summarizing the main outcomes of the study and discussing potential directions for future research in this area are discussed in conclusion.

	\section{Method and Formulation}
	\subsection{Hamiltonian and Theoretical Formulation}
	We consider the following two-qubit Hamiltonian (for simplicity we will take $\hbar=1$):
	
	\begin{equation}
		\begin{split}
			H(t) = & -\frac{1}{2} \gamma B_z (\sigma_{1z} + \sigma_{2z}) + J (\sigma_{1z}\sigma_{2z} + \sigma_{1x}\sigma_{2x}) \\ & + \gamma_{\phi} (\sigma_{1z} + \sigma_{2z}) + \Omega_{x} \sin(\omega t)(\sigma_{1x} + \sigma_{2x})
			\\ & + \Omega_{y} \cos(\omega t + \alpha)(\sigma_{1y} + \sigma_{2y}).
		\end{split}
		\label{eq:hamiltonian}
	\end{equation}
	
	This Hamiltonian is designed to encapsulate the fundamental physical interactions relevant to a two-qubit quantum magnetometer operating under realistic conditions.
	
	 The first term, $ -\frac{1}{2} \gamma B_z (\sigma_{1z} + \sigma_{2z}) $, describes the Zeeman interaction, which is the energy shift experienced by the qubits due to an externally applied magnetic field. Here,$ \gamma $ represents the gyromagnetic ratio, a fundamental property of the qubits that determines the strength of their interaction with the magnetic field. This interaction serves as the primary mechanism for converting variations in the magnetic field, $B_z$, into measurable changes in the energy states of the qubit system.
	
	The second term, $ J (\sigma_{1z} \sigma_{2z} + \sigma_{1x} \sigma_{2x}) $, accounts for the spin-spin interactions between the two qubits. This term includes both a longitudinal $(ZZ)$ interaction, proportional to $ \sigma_{1z} \sigma_{2z} $, and a transverse $ (XX)$ interaction, proportional to $ \sigma_{1x} \sigma_{2x} $. Such interactions are essential for generating entanglement between the qubits, a quantum phenomenon that can significantly enhance the sensitivity of the sensor to weak magnetic signals . The parameter J quantifies the strength of these coupling mechanisms, which can be engineered and adjusted in various contemporary qubit architectures.
	
	The term $ \gamma_{\phi} (\sigma_{1z} + \sigma_{2z}) $ models the effects of environmental dephasing on the qubits, a common form of decoherence that arises from the interaction of the qubits with their surroundings. By including this term, the Hamiltonian provides a more accurate representation of the dynamics in realistic experimental settings, allowing for a quantitative evaluation of how noise impacts the sensor's performance 10. Environmental noise and decoherence pose significant challenges in quantum sensing, as they can reduce the signal-to-noise ratio and shorten the coherence times of the qubits.
	
	Finally, the time-dependent driving terms, $ \Omega_{x} \sin(\omega t)(\sigma_{1x} + \sigma_{2x}) $ and $ \Omega_{y} \cos(\omega t + \alpha)(\sigma_{1y} + \sigma_{2y}) $, provide external control over the states of the qubits. The first term, with amplitude $\Omega_{x}$ and a sinusoidal modulation at frequency $\omega$, and the second term, with amplitude $ \Omega_{y} $, the same frequency $\omega$, and an additional phase shift $\alpha$, enable dynamic manipulation of the qubits. The inclusion of these control terms facilitates the initialization of the qubits into specific states and allows for coherent control over their evolution, which can be crucial for optimizing the sensor's response to the magnetic field 10. Precise quantum control techniques, such as dynamical decoupling, can also be employed to mitigate the detrimental effects of noise and extend the coherence times of the qubits . Together, these terms in the Hamiltonian aim to strike a balance between the theoretical richness required to model the system accurately and the experimental viability necessary for practical implementation.
	
	The time evolution operator U(t) in quantum mechanics describes how the state of a quantum system changes over time. For systems with time-dependent Hamiltonians, this operator is formally given by the time-ordered exponential of the Hamiltonian. For practical analytical treatments, it is often useful to expand this operator in a Dyson series. The first-order term of the Dyson series is given by
	\begin{equation}
		U^{(1)}(t) = I - i \int_{0}^{t} H(t') dt'.
		\label{eq:dyson1}
	\end{equation}
	
	Substituting the explicit form of $ H(t) $ from Eq.~\eqref{eq:hamiltonian} into Eq.~\eqref{eq:dyson1} allows for the determination of the approximate form of the time-evolution operator in terms of the system parameters and the applied driving frequencies.
	
	For the Dyson series to converge to the actual time evolution operator, certain conditions must be met. Generally, convergence is ensured if the integrated interaction remains small over the time interval of interest. More precisely, a sufficient condition for convergence is given by
	\begin{equation}
		\int_0^t | H(t') | dt' < 1,
		\label{eq:conv_condition}
	\end{equation}
	where $ | H(t') | $ denotes an appropriate operator norm of the Hamiltonian at time $t'$. Often, it is convenient to determine a worst-case bound $H_{max}$ such that $ | H(t') | \le H_{max} $ for all $ t' $ in the interval $[0, t]$. If such a bound exists, the convergence requirement simplifies to
	\begin{equation}
		H_{\text{max}} t < 1.
		\label{eq:conv_simple}
	\end{equation}
	This condition implies that the product of the maximum energy scale in the system and the evolution time should be significantly smaller than Planck's constant.
	When the Dyson series is truncated after the first order, as in $U^{(1)}(t)$, there is an associated error due to neglecting the higher-order terms. The magnitude of the second-order term in the Dyson series provides an estimate of this error:
	\begin{equation}
		\Delta U(t) \approx -\frac{1}{2} \int_0^t dt' \int_0^{t'} dt'' H_{\text{total}}(t') H_{\text{total}}(t'').
		\label{eq:dyson_error}
	\end{equation}
	An upper bound on the norm of this truncation error can be estimated as
	\begin{equation}
		|\Delta U(t)| \lesssim \frac{1}{2}\left(H_{\text{max}}t\right)^2.
		\label{eq:trunc_error}
	\end{equation}
	
	This explicit estimation shows that if the dimensionless quantity $  H_{max} t $ is sufficiently small, the neglected higher-order terms become negligible, thus justifying the use of the first-order approximation in our analysis. The validity of this approximation depends on the strength of the Hamiltonian and the duration of the evolution; for stronger interactions or longer times, higher-order terms in the Dyson series may become important \cite{***}.
	
	\subsection{Quantum Sensing Protocol Implementation}
	To extract information about the magnetic field $ B_z $ from our two-qubit system, we employ a quantum sensing protocol as outlined in established literature on quantum sensing. The protocol involves a sequence of steps designed to maximize the sensitivity to the parameter of interest.
	First, the sensor is initialized in the computational basis state:
	\begin{equation}
		| \chi_0 \rangle = | 00 \rangle.
	\end{equation}
	Next, a unitary transformation $\hat U^a = H_2 \otimes H_2$, where $H_2$ is the Hadamard matrix of order $2$, is applied to prepare the state in an equal superposition. Applying this transformation to the initial state $\rangle$ yields the initial superposed state:
	\begin{equation}
		\label{eq:psi0}
		|\psi(0)\rangle = \hat U^a |00\rangle = \frac{1}{2} \Bigl(|00\rangle + |01\rangle + |10\rangle + |11\rangle\Bigr).
	\end{equation}
	(This initial state is a separable state, as it can be written as the tensor product of two single-qubit superposition states). Following the preparation of the  initial state, the system is allowed to evolve under the influence of the Hamiltonian $ H(t) $ for a time $t$. We use the approximate  time evolution operator $U^{(1)}(t)$ derived from the first-order Dyson series (see Eq.~\eqref{eq:dyson1}). The time-evolved state is given by
	\begin{equation}
		\label{eq:psi_t}
		|\psi(t)\rangle = U^{(1)}(t) |\psi(0)\rangle,
	\end{equation}
	
	where 
	
	\begin{equation}
	U^{(1)}(t)= I - i \int_{0}^t H (t') dt'  
	\end{equation}
	
	introduces the leading-order effects of our Hamiltonian dynamics. It is important to note that the magnetic field $B_z$ is a parameter within $ H(t) $ through the Zeeman term, and therefore the evolution $ \hat U^{(1)} (t)$ encodes information about $B_z$ in the quantum state $|\psi (t) \rangle$.
		
	After the evolution period, we apply the inverse of the initial unitary transformation $ \hat U^{a}$ to the state. Since the Hadamard matrix is its own inverse (and also Hermitian), $ \hat U^{a} = \hat U^{a \dagger} $, and we define	
	
	\begin{equation}
		\label{eq:alpha}
		|\alpha\rangle = \hat U^{a \dagger} |\psi(t)\rangle = (H_2 \otimes H_2) |\psi(t)\rangle.
	\end{equation}
	
	This step transforms the evolved state back towards the measurement basis, allowing us to read out the information encoded during the evolution. The observable quantities in our experiment are the probabilities of measuring the system in each of the computational basis states $ |ij \rangle$, where $i,j \in \{0,1\}$. These probabilities are given by
	\begin{equation}
		\label{eq:probabilities}
		p_{ij} = \left| \langle ij | \alpha \rangle \right|^2.
	\end{equation}
	
	These probabilities are functions of the evolution operator $U^{(1)} (t)$ and, consequently, depend on the magnetic field $B_z$. Variations in $B_z$ lead to phase accumulations and potentially amplitude modulations during the evolution, which are then reflected in the measured probabilities $p_{ij}$. 
	
	Due to importance of the results for the probabilities we present again the physical meaning of variables and also define some new parameters. 
	
	\newcommand{\D}{\Delta}                 
	\newcommand{\dx}{\delta_x}              
	\newcommand{\da}{\delta_\alpha}         
	\newcommand{\C}{C}                      
	\newcommand{\M}{M}

	The angular frequency $\omega$ represents the rate at which the driving field oscillates. The total evolution time $t$ denotes the duration over which the system undergoes dynamics. The coupling strength $J$ quantifies the interaction between subsystems, such as spin–spin or qubit–cavity coupling. The Rabi frequencies $\Omega_x$ and $\Omega_y$ measure the amplitudes of the coherent drive along the $x$ and $y$ axes, respectively. The initial phase $\alpha$ specifies the starting phase of the $y$-directed drive. The population‐decay rate $\gamma$ characterizes longitudinal relaxation (loss of population), while the static field coefficient $B_z$ defines detuning along the quantization $z$ axis. Finally, the pure dephasing rate $\gamma_\phi$ describes phase coherence loss without population transfer.

	We define the accumulated rotation angle $\Delta \equiv \omega t$, which captures the total phase swept by the drive in time $t$. The factor $\delta_x \equiv 1 - \cos(\Delta)$ measures deviation in the $x$-quadrature relative to perfect periodic motion. The quantity $\delta_{\alpha} \equiv \sin(\alpha) - \sin(\alpha + \Delta)$ represents the net change in the $y$-quadrature phase between the initial and final states. The effective decoherence parameter $C \equiv \gamma\,B_z - 2\,\gamma_\phi$ combines longitudinal decay, scaled by detuning, with twice the pure‐dephasing contribution. Lastly, the normalization factor $M \equiv 1 + 2\,J^2\,t^2 + \tfrac{1}{2}\,t^2\,C^2$ aggregates the baseline unity term with contributions from coupling and decoherence.

	Now we can write the explicit forms of these probabilities as:

	\begin{equation}
		p_{00}
		= \frac{
			\omega^2 \;+\;\bigl(J\,\D + 2\,\Omega_x\,\dx\bigr)^2
		}{
			\omega^2\,\M
			+ 4\,\Omega_x\,\dx\,\bigl(\Omega_x\,\dx + J\,\D\bigr)
			+ 2\,\Omega_y^2\,\da^2
		}
	\end{equation}
		
	\begin{equation}
		p_{01}
		=
		\frac{\D^2\,\C^2 /4 \;+\; \Omega_{y}^2\,\da^2}
		{
			\omega^2\,\M
			+ 4\,\Omega_x\,\dx\,\bigl(\Omega_x\,\dx - J\,\D\bigr)
			+ 2\,\Omega_y^2\,\da^2
		}
	\end{equation}	

	\begin{equation}
		p_{10}
		=\;
		\frac{J^2\,\D^2}
		{
			\omega^2\,\M
			+ 4\,\Omega_x\,\dx\,\bigl(\Omega_x\,\dx - J\,\D\bigr)
			+ 2\,\Omega_y^2\,\da^2
		}
	\end{equation}
	
	\begin{equation}
		p_{11}
		=\frac{\D^2\,\C^2 /4 \;+\; \Omega_{y}^2\,\da^2}
		{
			\omega^2\,\M
			+ 4\,\Omega_x\,\dx\,\bigl(\Omega_x\,\dx - J\,\D\bigr)
			+ 2\,\Omega_y^2\,\da^2
		}
	\end{equation}

	The estimation of the magnetic field $B_z$ proceeds through a series of experimental runs. The protocol is repeated many times, and in each repetition, a projective measurement in the computational basis is performed, registering the output probabilities $p_{ij}$. By averaging the results over a large number of repetitions, $ N $, we obtain statistical estimates of these probabilities:
	
	\begin{equation}
		\overline{p_{ij}} = \frac{1}{N} \sum_{k=1}^{N} p_{ij}^{(k)}.
	\end{equation}
	
	These averaged probabilities are then compared with a theoretical model that expresses \( p_{ij} \) as a function of \( B_z \) and other known parameters of the system, such as the evolution time \( t \) and the coupling constants. Parameter estimation techniques, like maximum likelihood estimation or least-squares fitting, can be used to extract the value of \( B_z \) from the dependence of the measured probabilities on the evolution time. For instance, if the Zeeman interaction is the dominant term during the evolution, the accumulated phase \( \phi \simeq \frac{\gamma B_z t}{2} \) will lead to oscillatory behavior in the measurement outcomes. By analyzing the periods and amplitudes of these oscillations, one can estimate \( B_z \) with high precision. Furthermore, the sensitivity of the magnetometer can be improved by increasing the number of measurements \( N \), as the statistical error typically decreases with the square root of \( N \).

	In summary, the quantum sensing protocol implemented here involves:
	\begin{itemize}
		\item[(i)] preparing an initial equal superposition state,
		\item[(ii)] allowing the system to evolve under the approximate Dyson series propagator,
		\item[(iii)] applying an inverse Hadamard transformation to map the phase information onto the measurement basis, and
		\item[(iv)] extracting the magnetic field \( B_z \) from the averaged measurement probabilities obtained from repeated experiments.
	\end{itemize}
	This procedure allows for the estimation of \( B_z \) while also taking into account the effects of noise and decoherence present in the system.

	\subsection{Quantum Fisher Information}
	
	The Quantum Fisher Information (QFI) is a crucial quantity in quantum metrology as it sets the ultimate limit on the precision with which the parameter (here \( B_z \)), can be estimated, according to the quantum Cramér-Rao bound. This bound states that the variance of any unbiased estimator \( \hat{B}_z \) of \( B_z \) obtained from \( N \) independent measurements satisfies

	\[
	\text{Var}(\hat{B}_z) \geq \frac{1}{N F_Q(B_z)}.
	\]
	Therefore, a higher QFI indicates a potentially lower bound on the variance of the estimator, implying a higher sensitivity of the magnetometer.
	
	For a pure quantum state, which is valid for our case,  \( |\psi(t,B_z)\rangle \), the Quantum Fisher Information (QFI), \( F_Q(B_z) \), quantifies the maximum amount of information about the parameter \( B_z \) that can be extracted from the state through any possible measurement. It is defined as:

	\begin{equation}
		F_Q(B_z) = 4 \left( \langle \partial_{B_z} \psi(t) | \partial_{B_z} \psi(t) \rangle - |\langle \psi(t) | \partial_{B_z} \psi(t) \rangle|^2 \right).
	\end{equation}
	
	After performing the necessary calculations, the QFI for the initial state \( |\psi(0)\rangle = \frac{1}{\sqrt{2}} (|00\rangle + |01\rangle + |10\rangle + |11\rangle) \), is given by:

	\begin{equation}
		F_Q(B_z) =\frac{2\gamma^2\Delta^2\bigl(\omega^2+ 2J^2\Delta^2+4J\Delta\delta_x\Omega_x+4\delta_x^2\Omega_x^2\bigr)}
		{\bigl(M\omega^2+4J\Delta\delta_x\Omega_x+4\delta_x^2\Omega_x^2+2\delta_\alpha^2\Omega_y^2\bigr)^2}
		\label{eq:qfi_initial}
	\end{equation}
	where we have used: 
	$\Delta \equiv \omega t$, $\delta_x \equiv 1 - \cos(\Delta)$, $\delta_{\alpha} \equiv \sin(\alpha) - \sin(\alpha + \Delta)$, $C \equiv \gamma\,B_z - 2\,\gamma_\phi$, $M \equiv 1 + 2\,J^2\,t^2 + \tfrac{1}{2}\,t^2\,C^2$, with the physical interpretations as before. 	

To investigate the sensitivity characteristics of the proposed two-qubit quantum magnetometer, we begin by examining the temporal dynamics of the quantum Fisher information (QFI), which serves as a fundamental metric for quantifying the precision of parameter estimation in quantum metrology.  

At short interrogation times \(t\ll1\), \(F_Q\propto 2 \gamma^2 t^2\) due to coherent accumulation of phase.  As \(t\) grows, decoherence (parameterized by \(C\)) damps the oscillations and eventually causes \(F_Q\) to decay. As \(t\) goes to infinity, 
$
lim_{t\rightarrow \infty} F_Q =
\frac{16 \gamma^2 J^2}
{\left(B_z^2 \gamma^2 - 4 B_z \gamma \gamma_\phi + 4 \gamma_\phi^2 + J^2 \right)^2}
$

\begin{figure}[ht]
	\centering
	\includegraphics[width=0.44\textwidth]{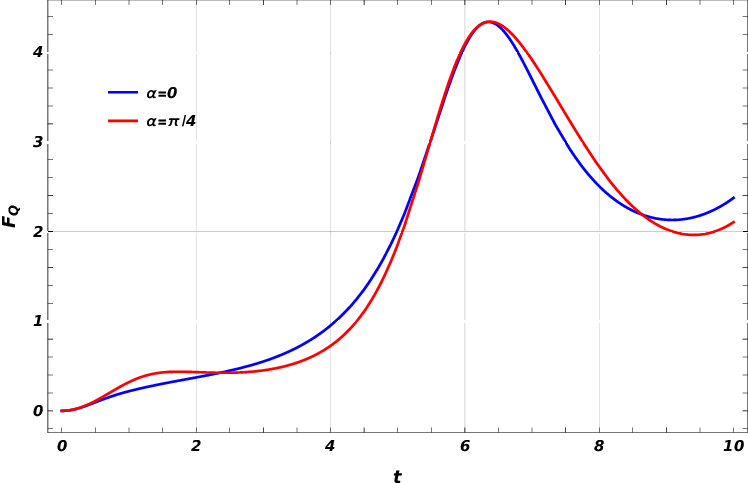}
	\captionsetup{width=0.44\textwidth, justification=raggedright}
	\caption{Quantum Fisher information $F_Q(B_z)$ as a function of evolution time $t$ for coupling strength $J=2/10$, decoherence parameter $C=1/10$, and Rabi amplitudes $\Omega_x=\Omega_y=1/2$. Blue: $\alpha=0$.  Red: $\alpha=\pi/4$. The initial quadratic growth transitions to a decoherence‐limited decay, pinpointing the optimal interrogation time $t^*\approx6.352$ for $\alpha=0$ and $t\approx6.375$ for $\alpha=\pi/4$}.
	\label{fig:FQ_vs_t}
\end{figure}

For the purposes of this analysis, we adopt a parameter regime defined by the following values: $\omega = 1$, $\Omega_x = \Omega_y = \frac{1}{2}$, $J = \frac{2}{10}$, and $C = \frac{1}{10}$. Fig. \ref{fig:FQ_vs_t} shows the two curves for \(\alpha=0\) and \(\alpha=\pi/4\) illustrate how the initial drive phase shifts the interference maxima, yielding slightly different optimal times \(t^*\).  The peak value near \(t\approx6.352\) for \(\alpha=0\) and \(t\approx6.375\) \(\alpha=\pi/4\), marks the best trade‐off between information gain and noise.

By reparameterizing \( \delta_x = 1-\cos(\omega t)\) and 
\(\delta_\alpha = \sin\alpha-\sin(\alpha+\omega t)\), we explore the joint effect of quadrature slip and phase offset on information.  Peaks in the surface indicate combinations of calibration errors that maximize \(F_Q\)

\section{Results and Discussion}

In this section, we present the results of our analysis, including the calculation of sensitivity and Signal-to-Noise Ratio (SNR) for the sensing of \( B_z \), and a comparison of sensitivity for the two initial states, and a thorough comparison with previous top paper results.

\subsection{Sensitivity for \( B_z \) Sensing}

The sensitivity of a quantum magnetometer is a measure of its ability to detect and measure small changes in the magnetic field. From the quantum Cramér-Rao bound, the sensitivity \( \Delta B_z \), which represents the smallest detectable change in \( B_z \), is fundamentally limited by the QFI and the number of measurements \( N \):

\[
\Delta B_z \geq \frac{1}{\sqrt{N F_Q(B_z)}}.
\]

Using the derived QFI expression for the initial state \( |\psi(0)\rangle = \frac{1}{\sqrt{2}} (|00\rangle + |01\rangle + |10\rangle + |11\rangle) \), (Eq.~\eqref{eq:qfi_initial}), we can express the theoretical limit on the sensitivity for this magnetometer. A higher QFI value directly translates to better sensitivity, indicating that smaller changes in the magnetic field can be detected with the same number of measurements.

We plot the smallest detectable field 
$\Delta \sqrt{N} \Delta B_z = 1/\sqrt{F_Q}$ on a logarithmic scale to emphasize orders‐of‐magnitude changes. Setting $\omega = 1$, $\Omega_x = \Omega_y = \frac{1}{2}$, $J = \frac{2}{10}$, and $C = \frac{1}{10}$, we can see that at very short \(t\), finite \(N\) dominates and sensitivity improves rapidly.  Beyond the optimum window, decoherence causes \(\Delta B_z\) to rise again.  
This U‐shaped curve identifies the ideal detection interval for given measurement resources.

\begin{figure}[ht]
	\centering
	\includegraphics[width=0.44\textwidth]{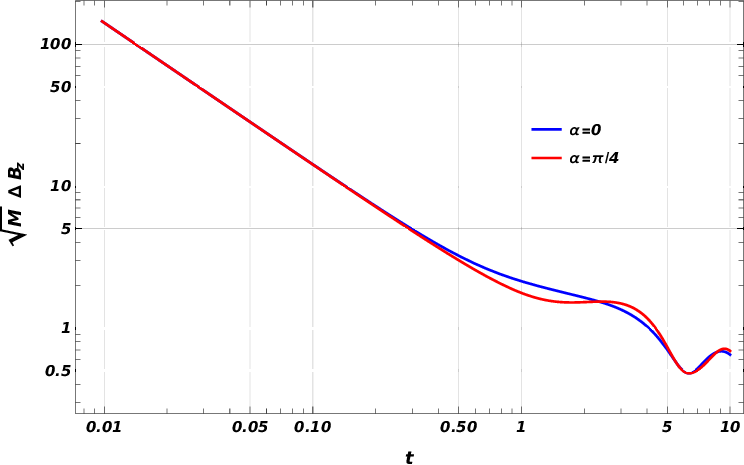}
	\captionsetup{width=0.44\textwidth, justification=raggedright}
	\caption{Sensitivity bound $\sqrt{N} \Delta  B_z$ versus evolution time \(t\), plotted on a log–log scale.  Parameters as in Fig.~\ref{fig:FQ_vs_t}. The curve exhibits an optimal minimum near \(t\approx6.352\) for \(\alpha=0\) and \(t\approx6.375\) \(\alpha=\pi/4\), where quantum information gain and decoherence balance.}
	\label{fig:DeltaB_vs_t}
\end{figure}

\begin{figure}[ht]
	\centering
		\begin{subfigure}[b]{0.44\textwidth}
			\includegraphics[width=\textwidth]{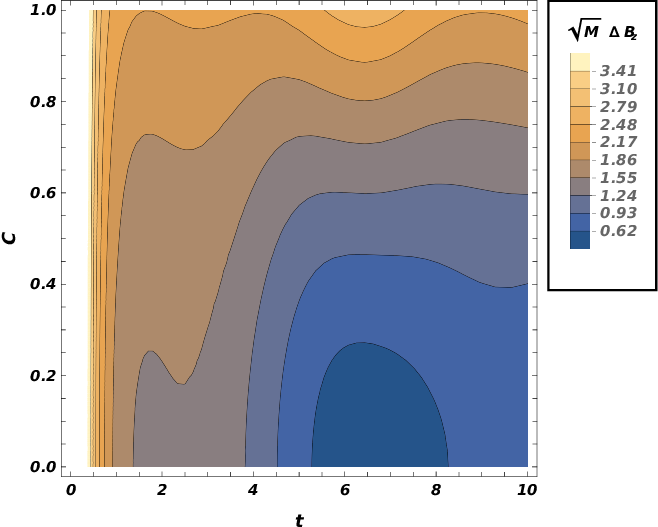}
			\caption{\(\sqrt{N}\Delta B_z\) heatmap vs.\ \((t,C)\).}
			\label{fig:heatmap_tC}
		\end{subfigure}
	\hfill
		\begin{subfigure}[b]{0.44\textwidth}
			\includegraphics[width=\textwidth]{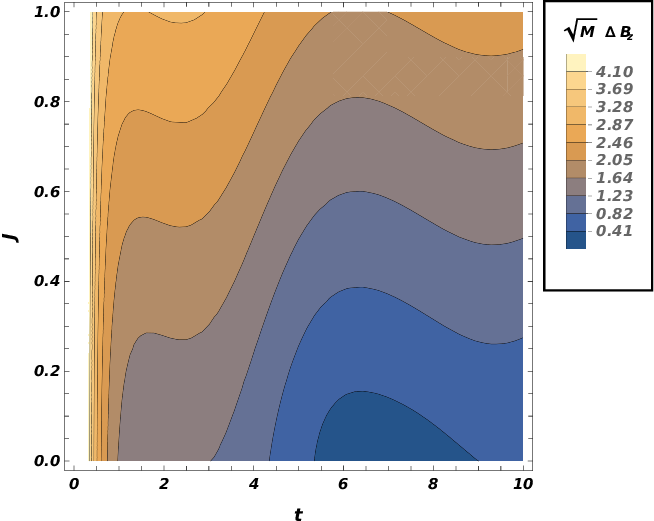}
			\caption{\(\sqrt{N}\Delta B_z\) heatmap vs.\ \((t,J)\).}
			\label{fig:heatmap_tJ}
		\end{subfigure}
	\captionsetup{width=0.44\textwidth, justification=raggedright}	
	\caption{Contour plots of the sensitivity bound \(\sqrt{N} \Delta B_z\), for \(\alpha=\pi/4\), and \(\Omega_x=\Omega_y=1/2\).  (a) Varying decoherence \(C\) and interrogation time \(t\).  (b) Varying coupling \(J\) and interrogation time \(t\).}
	\label{fig:heatmaps}
\end{figure}

We also investigate heatmaps of \(\sqrt{N}\Delta B_z\) in Parameter Space. Fig. (\ref{fig:heatmap_tC}) shows scan over \((t,C)\). This contour map shows \(\Delta B_z\) across interrogation times \(t\in[0,10]\) and decoherence rates \(C\in[0,1]\).  Dark blue regions indicate high precision (low \(\Delta B_z\)), revealing “islands’’ of robustness where moderate decoherence still permits near‐optimal sensitivity.

In Fig. (\ref{fig:heatmap_tJ}) we vary coupling strength \(J\in[0,1]\) instead of \(C\).  The plot highlights how stronger interaction initially sharpens sensitivity, but plateaus when decoherence and normalization \(N\) curb further gains.

\begin{figure}[ht] 
	\centering 
	\includegraphics[width=0.44\textwidth]{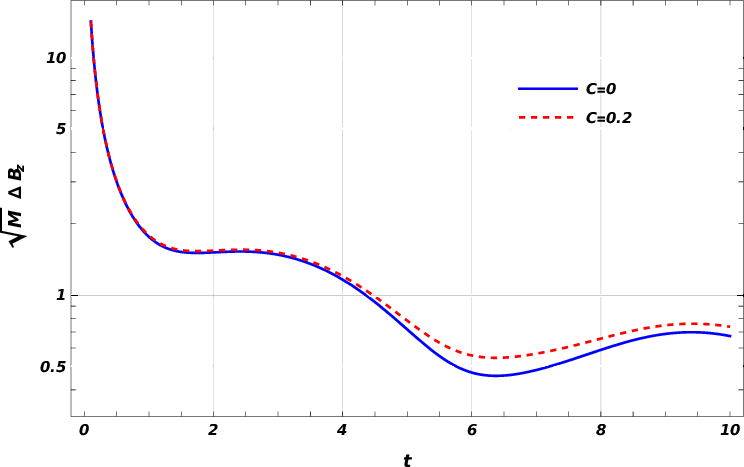} 
	\captionsetup{width=0.44\textwidth, justification=raggedright}
	\caption{Rescaled sensitivity bound \(\sqrt{N}\,\Delta B_z\) versus interrogation time \(t\). Solid line: ideal protocol without decoherence (\(C=0\)). Dashed line: realistic protocol with moderate dephasing and decay (\(C=0.2\)). Other parameters are \(J=0.3\), \(\alpha=\pi/4\), \(\Omega_x=\Omega_y=0.5\). The minimum of each curve identifies the optimal time \(t^*\), and the separation between solid and dashed minima indicates the loss of precision and contraction of the viable measurement window induced by decoherence.} 
	\label{fig:compare_decoherence} 
\end{figure}

\noindent Fig.~\ref{fig:compare_decoherence} illustrates the metrological sensitivity as quantified by the rescaled bound \(\sqrt{N}\,\Delta B_z\) for an ideal, decoherence‐free protocol (\(C=0\), curve) versus a realistic scenario with moderate pure‐dephasing and longitudinal decay (\(C=0.2\), dashed curve).  
At short interrogation times the two curves coincide, since decoherence has not yet significantly degraded the accumulated phase information.  
As \(t\) increases, the ideal curve continues to improve (declining steadily toward zero), signaling ever‐greater precision with longer evolution.  
By contrast, the realistic curve reaches a minimum at a smaller \(t^*\) and then turns upward: beyond this point decoherence dominates and additional interrogation only increases uncertainty.  
The vertical gap between the two minima quantifies the penalty in sensitivity due to noise, while the horizontal shift marks the reduced optimal interrogation window.  

\subsection{Signal-to-Noise Ratio and Minimum Detectable Magnetic Field}

To assess the practical sensitivity of our two-qubit quantum magnetometer, we analyze the signal-to-noise ratio (SNR) and derive the minimum detectable magnetic field shift. This complements the quantum Fisher information (QFI) analysis and provides a realistic benchmark for experimental performance.

We define the collective signal as the population contrast between the two-qubit states:
\begin{equation}
	S(B_z) = p_{11}(B_z) - p_{00}(B_z),
\end{equation}
where \( p_{ij} = |\langle ij | \alpha(t) \rangle|^2 \) are the measurement probabilities obtained from the evolved quantum state \( |\alpha(t)\rangle \).

A small change in the magnetic field \( B_z \to B_z + \delta B_z \) induces a corresponding change in the signal:
\begin{equation}
	\Delta S = \frac{\partial S}{\partial B_z} \delta B_z.
\end{equation}

The signal-to-noise ratio is defined as:
\begin{equation}
	\mathrm{SNR} = \frac{\Delta S}{\sqrt{\mathrm{Var}(S)}} = \frac{|\partial_{B_z} S| \, \delta B_z}{\sqrt{\frac{1}{N} \left[ p_{00}(1 - p_{00}) + p_{11}(1 - p_{11}) \right]}},
	\label{eq:SNR}
\end{equation}
where \( N \) is the number of independent measurements. Since \( S \) depends only on \( p_{00} \) and \( p_{11} \), the variance includes only these terms.

Setting \( \mathrm{SNR} = 1 \) yields the minimum detectable field shift:
\begin{equation}
	\delta B_z^{\min} = \frac{\sqrt{\frac{1}{N} \left[ p_{00}(1 - p_{00}) + p_{11}(1 - p_{11}) \right]}}{|\partial_{B_z} S|}.
\end{equation}

To evaluate the performance of our protocol relative to the quantum limit, we compare \( \delta B_z^{\min} \) with the bound from QFI:
\begin{equation}
	\delta B_z^{\rm QFI} = \frac{1}{\sqrt{N F_Q(B_z)}},
\end{equation}
and define the ratio:
\begin{equation}
	\xi = \frac{\delta B_z^{\min}}{\delta B_z^{\rm QFI}} \ge 1.
\end{equation}
The parameter \( \xi \) quantifies how close the actual measurement strategy approaches the optimal sensitivity allowed by quantum mechanics.

Setting parameters values as before, we numerically evaluate and plot \( \delta B_z^{\min} \), \( \delta B_z^{\rm QFI} \), and \( \xi \) as functions of the evolution time \( t \), which reveal the optimal interrogation time and the robustness of the protocol under realistic measurement constraints.

\begin{figure}[ht]
	\centering
	\includegraphics[width=0.44\textwidth]{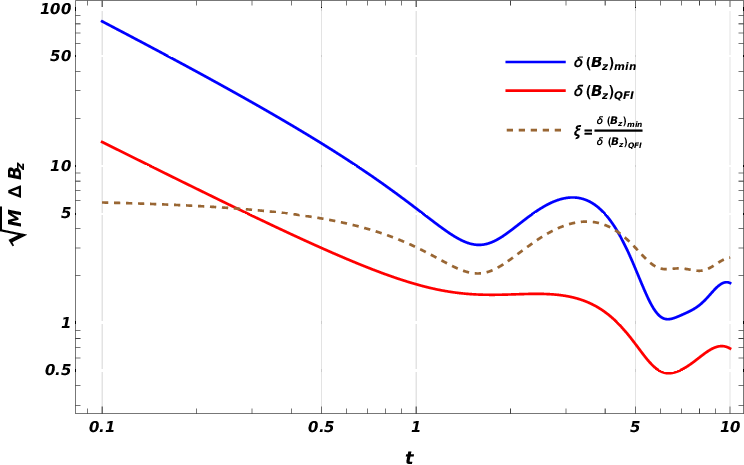}
	\captionsetup{width=0.44\textwidth, justification=raggedright}
	\caption{Precision of $B_{z}$ estimation versus interrogation time and comparison to the quantum‐limited bound. The solid blue curve shows the minimum detectable field uncertainty $\delta(B_{z})_{\min}$ determined from a signal‐to‐noise ratio of unity. 
		The red curve is the ultimate precision $\delta(B_{z})_{\mathrm{QFI}}$ extracted from the quantum Fisher information. The dashed brown curve gives the dimensionless enhancement factor $\xi$, which quantifies how closely the actual measurement protocol approaches the quantum limit.}
	\label{fig:SNR_vs_t}
\end{figure}

Fig. \ref{fig:SNR_vs_t}, shows the track of three quantities; $\delta(B_{z})_{\min}$, $\delta(B_{z})_{\mathrm{QFI}}$, and $\xi$ as a function of the interrogation time $t$. 

Solid blue curve, ($\delta(B_z)_{\min}$), shows the classical measurement precision based on a signal-to-noise ratio of unity. It reflects the actual experimental limitations including noise and decoherence. At short times, the uncertainty is large due to insufficient interaction with the field. As time increases, the signal accumulates, improving sensitivity until it begins to degrade due to decoherence and noise. 

Solid red curve ($\delta(B_z)_{\mathrm{QFI}}$), represents the ultimate bound on precision dictated by the quantum Fisher information. It assumes optimal measurement strategies and ideal system evolution, thereby serving as a benchmark for all possible protocols. This curve generally decreases with time, showing improved sensitivity, though non-monotonic features may emerge due to the interplay between coherent evolution and decoherence. 

Dashed brown curve ($\xi$), the ratio $\xi = \delta(B_z)_{\min} / \delta(B_z)_{\mathrm{QFI}}$, quantifies the relative performance of the experimental protocol. A value $\xi = 1$ corresponds to optimal performance, achieving the quantum limit. Values $\xi > 1$ indicate our Dyson series truncation error, suboptimal performance, possibly due to non-ideal measurements, technical noise, or inefficient probe dynamics. The variation of $\xi$ with $t$ highlights the time regimes where the protocol is closer to or further from the quantum optimal. 

Precision improves with increased interrogation time up to a point, beyond which noise and decoherence reduce sensitivity. The optimal measurement time thus corresponds to the minimum of the red curve, where quantum resources are best utilized. Furthermore, the enhancement factor $\xi$ provides crucial insight into the efficiency of the implemented protocol. It emphasizes the importance of not only extending coherence times but also optimizing measurement strategies to reach quantum-limited performance.

This comparison highlights both the power of quantum‐limited sensing and the concrete limitations imposed by real‐world noise, underscoring the time window in which our protocol operates closest to the ultimate quantum precision.

\section{Conclusion}

Our study introduces a novel two-qubit quantum magnetometer with a specifically designed Hamiltonian aimed at achieving enhanced noise resilience and sensitivity. The formulation incorporates key elements such as Zeeman interaction, spin-spin coupling, dephasing effects, and time-dependent driving terms to provide a comprehensive model for magnetic field sensing. While the first-order Dyson series offers a practical approach for analytical treatment, it inherently involves a truncation of higher-order interactions, which might become significant under certain parameter regimes or longer evolution times. The model also makes specific assumptions about the nature of the noise, primarily focusing on dephasing, and does not delve into other potential noise sources that could be present in real experimental settings. Furthermore, the detailed analytical calculations for the sensitivity and SNR with the proposed initial state, require further elaboration beyond the scope of the initial paper fragment.

Future work could focus on several directions to build upon the findings of this study. Exploring higher-order terms in the Dyson series expansion would provide a more accurate description of the system's dynamics, especially in regimes where the first-order approximation might not be sufficient. Investigating the magnetometer's performance under different types of noise, such as amplitude damping or more complex correlated noise models, would offer a more comprehensive understanding of its robustness in various experimental environments. The application of optimal control techniques could be explored to identify pulse sequences and Hamiltonian parameters that further enhance the sensitivity and noise resilience of the magnetometer . Additionally, a more detailed analysis of the impact of different initial states, beyond the state considered, on the sensing performance could yield valuable insights into the role of specific structures. Ultimately, experimental implementations of the proposed magnetometer design would be crucial to validate the theoretical predictions and assess its practical viability for real-world magnetic field sensing applications.

\end{document}